\documentclass[12pt]{article}
\usepackage{epsfig}
\usepackage{amssymb}
\usepackage{amsmath,amsfonts,amssymb,graphicx}
\newcommand {\be}{\begin{equation}}
\newcommand {\ee}{\end {equation}}
\newcommand{\beq}{\begin{eqnarray}}
\newcommand{\eeq}{\end{eqnarray}}
\begin{document}
\title{Erratum for Time Reversal in Neutrino Oscillations}
\author{Leonard S. Kisslinger$^a$, Ernest M. Henley$^{b}$, and 
Mikkel B. Johnson$^{c}$\\ 
 $^a$Department of Physics, Carnegie Mellon University, Pittsburgh, 
PA 15213\\
 $^b$Department of Physics, University of Washington, Seattle,
WA 98195 \\
 $^c$Los Alamos National Laboratory, Los Alamos, NM 87545}
\date{}
\maketitle

In our recent work\cite{hjk11} we estimated time reversal violations (TRV)
for a number of projects that detect neutrino oscillations. Since none
of these facilities have both electron and muon neutrino beams, tests of
TRV are not possible. We suggested a new experiment for the MINOS\cite{minos}
project, with baseline L=735 km, to employ another detector at baseline 2L,
which could measure TRV, as with a 10\% e-mu conversion at distance L, there
would be a 1\% increase of the original flavor at 2L, which could be detected. 
However, we neglected the curvature of the earth. An essential part of the
calculation in Ref\cite{hjk11} is the interaction V as the neutrino beam
moving through earth interacts with electron. As a result, V=0 for the
L to 2L distance, as shown in Fig. 1

If a detector were to be placed at 1470 km, as in the proposed L-2L experiment,
it would have to be installed about 1.3 km above ground, position 2.
Our new proposed experiment is shown in Fig. 1, in which the neutrino beam
is aimed at a new detector at the surface of the earth 735 km past the
Soudan mine (position 3 in the figure), only a small deviation from the
present MINOS neutrino beam. Since position 3 is very similar to that of
the MINOS near detector at FNAL\cite{minos}, this detector would be much 
easier and less expensive to construct than that proposed in Ref\cite{hjk11}.
There would be a 1\% increase in the electron neutrino probability that 
one would obtain with the 10\% muon neutrino flux at the position of the 
Soudan mine. We conclude that this would be a means for the measurement of 
time reversal violation with neutrino oscillations.

\begin{figure}[ht]
\begin{center}
\epsfig{file=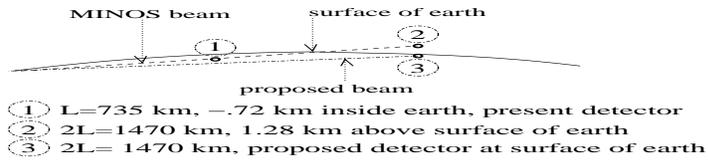,height=2.0cm,width=10cm}
\end{center}
\caption{The present Minos neutrino beam and a proposed beam}
\end{figure}

\Large{{\bf Acknowledgements}}\\
\normalsize
This work was supported in part by a grant from the Pittsburgh Foundation,
and in part by the DOE contracts W-7405-ENG-36 and DE-FG02-97ER41014. We
also acknowledge helpful discussions with Dr. William Louis, LANL.

\end{document}